\RequirePackage{fix-cm}
\documentclass[smallextended, final]{svjour3} 
\usepackage[english]{babel}
\usepackage[T1]{fontenc}
\usepackage{graphicx}
\usepackage{amssymb, amsmath}
\usepackage{cite}
\usepackage{fixltx2e}
\usepackage{fullpage}

\newcommand{\Kn}{\mathrm{Kn}}
\newcommand{\dd}{\mathrm{d}}
\newcommand{\pder}[2][]{\frac{\partial#1}{\partial#2}}
\newcommand{\pderdual}[2][]{\frac{\partial^2#1}{\partial#2^2}}
\newcommand{\pderder}[3][]{\frac{\partial^2#1}{\partial#2\partial#3}}
\newcommand{\Pder}[2][]{\partial#1/\partial#2}

\journalname{Theoretical and Computational Fluid Dynamics}

\begin{document}

\title{
    Computer simulation of slightly rarefied gas flows driven by significant temperature variations and their continuum limit
}

\author{O.A.~Rogozin}
\institute{O.A.~Rogozin \at
    Moscow Institute of Physics and Technology,
    9 Institutskiy pereulok, g. Dolgoprudny,
    Moskovskaya obl., Russian Federation\\
    \email{oleg.rogozin@phystech.edu}
}

\date{Received: date / Accepted: date}

\maketitle

\begin{abstract}
    A rigorous asymptotic analysis of the Boltzmann equation for small Knudsen numbers
    leads, in the general case, to more complicated sets of differential equations
    than widely used to describe the behavior of gas in terms of classical fluid dynamics.
    The present paper deals with such one that is valid for significant temperature variations
    and finite Reynolds numbers at the same time (slow nonisothermal flow equations).
    A finite-volume solver developed on the open-source CFD platform OpenFOAM\textregistered{}
    is proposed for computer simulation of a slightly rarefied gas in an arbitrary geometry.
    Typical temperature driven flows are considered as numerical examples.
    A force acting on uniformly heated bodies is studied as well.
\end{abstract}

\section{Introduction}

The Navier--Stokes set of equations are usually used for modeling of a gas with the vanishing mean free path.
Slightly rarefied gas effects, which are widely thought to take place only in the thin Knudsen layer,
are taken into account using special slip conditions on the boundary~\cite{SharipovCoefficients}.
In some situations, however, the Navier--Stokes set fails to describe the correct density
and temperature fields even in the continuum limit~\cite{Kogan1976}.

When the Reynolds number is finite as well as the temperature variations:
\[ \mathrm{Re} = O(1), \quad \frac1T\left|\pder[T]{x_i}\right| = O(1), \]
higher-order terms of gas rarefaction begin to play a significant role in the gas behavior.
Following~\cite{Kogan1976}, we will call such flows as~\emph{slow nonisothermal}.
In this case, the asymptotic analysis of the Boltzmann equation arrives on the scene and
yields the appropriate equations on the macroscopic variables, together with
their associated boundary conditions~\cite{Sone2002, Sone2007}.

This advanced set of equations reveals that some flows of the first order of the Knudsen number
are caused in the gas due to temperature variations in the absence of the external force:
thermal creep and nonlinear thermal-stress. The former is a well-known flow induced along a boundary
with a nonuniform temperature~\cite{Maxwell1879Stresses, Ohwada1989Creep}.
The latter occurs in the gas where the distance between isothermal surfaces varies along them~\cite{Kogan1976}.
It is described first in terms of the Chapman--Enskog expansion~\cite{Kogan1971}.
Later works are based on the Hilbert expansion~\cite{SoneBobylev96}.
The nonlinear thermal-stress flow can be considered as another type of convection
that occurs in the state of weightlessness.
The nonlinear origin of this phenomenon is relevant, because a linear thermal-stress flow
appears only in the next order of the Knudsen number~\cite{Sone1972Stress}.
Detailed experimental study of the nonlinear thermal-stress flow is an intractable problem,
but some attempts are made in~\cite{ExperimentsNTFS2003}.

From the computational point of view, the new set of equations has the similar structure
to the Navier--Stokes one, so similar numerical methods can be utilized too.
The pioneering solutions are based on the finite-difference
methods~\cite{SoneBobylev96, Aleksandrov2002Tube, Aleksandrov2008Particle}.
The finite-volume method is applied in some recent works~\cite{Laneryd2006, Laneryd2007}.

Currently, many CFD platforms provide comprehensive
capabilities for numerical simulation of the Navier--Stokes equations,
but the considered fluid-dynamic-type set of equations remains in the shadow.
The goal of the present paper is to improve this situation and
draw more attention to such CFD scope extensions.

One of the modern and promising CFD platforms, OpenFOAM\textregistered{},
has been selected as a basis for numerical algorithms developing.
OpenFOAM\textregistered{} is an object-oriented C++ library of classes and routines for parallel computation,
providing a set of high-level tools for writing advanced CFD code~\cite{OpenFOAM1998}.
It has a wide set of basic features, similar to any commercial one~\cite{OpenFOAM2010},
and is a robust software widely used in the industry~\cite{BoilingFlows2009,
TurbulentCombustion2011, CoastalEngineering2013, BiomassPyrolysis2013}.
As for rarefied gas, OpenFOAM\textregistered{} has a standard solver for DSMC, which
can be extended for hybrid simulations~\cite{HybridSolver2014}.
Finally, the most important advantage of OpenFOAM\textregistered{} is its open-source code,
so it is easy to add any modification to any part of the implementation.
OpenFOAM\textregistered{} is also well documented and has a large and active community of users.

\section{Basic equations}

The behavior of a gas is governed by the conservation equations of mass, momentum, and energy:
\begin{gather}
    \pder[\rho]{t} + \pder{x_i}(\rho v_i) = 0, \label{eq:mass}\\
    \pder{t}(\rho v_i) + \pder{x_j}(\rho v_i v_j + p_{ij}) = \rho F_i, \label{eq:momentum}\\
    \pder{t}\left[\rho\left(e+\frac{v_i^2}2\right)\right] +
        \pder{x_j}\left[\rho v_j\left(e+\frac{v_i^2}2\right)+v_i p_{ij}+q_j\right] = \rho v_j F_j. \label{eq:energy}
\end{gather}
Macroscopic variables have the following notation: the density \(\rho\), the flow velocity \(v_i\),
the stress tensor \(p_{ij}\), the specific internal energy \(e\), and the heat-flow vector \(q_i\).
Finally, \(F_i\) denotes the external force.
For an ideal monatomic gas, the internal energy \(e = 3RT/2\) depends only on the temperature;
\(R = k_B / m\) is the specific gas constant.
The pressure is taken from the equation of state \( p = \rho RT \).

The Navier--Stokes set equations is obtained from the conservation equations with the help of Newton's law
\begin{equation}\label{eq:Newton_law}
    p_{ij} = p\delta_{ij} - \mu\left(\pder[v_i]{x_j}+\pder[v_j]{x_i}-\frac23\pder[v_k]{x_k}\delta_{ij}\right) -
        \mu_B\pder[v_k]{x_k}\delta_{ij},
\end{equation}
and Fourier's law
\begin{equation}\label{eq:Fourier_law}
    q_i = -\lambda\pder[T]{x_i}.
\end{equation}
Some of the transport coefficients are used here:
the viscosity \(\mu\), the bulk viscosity \(\mu_B\), and the thermal conductivity \(\lambda\).
The bulk viscosity of an ideal monoatomic gas is zero (\(\mu_B = 0\)).

The Knudsen number \(\Kn = \ell/L\), the parameter characterizing the rate of rarefaction of the gas,
is determined by the ratio of the mean free path \[ \ell = \frac{m}{\sqrt2\pi d_m^2 \rho} \]
to the reference length \(L\).
For a hard-sphere gas, the radius of the influence range of the intermolecular force \(d_m\)
coincides with the diameter of a molecule.
The viscosity \(\mu\) and the thermal conductivity \(\lambda\) of an ideal gas
are proportional to the mean free path \(\ell\) and therefore to the Knudsen number:
\begin{equation}
    \mu = O(\Kn), \quad \lambda = O(\Kn).
\end{equation}

In the continuum limit (\(\Kn\to0\)), we obtain the Euler set of equations:
\begin{equation}
    p_{ij} = p\delta_{ij}, \quad q_i = 0,
\end{equation}
which makes it impossible to determine the temperature field of a gas at rest.
The classical heat-conduction equation is derived from~\eqref{eq:energy} and~\eqref{eq:Fourier_law}
in the absence of gas flows (\(v_i = 0\)):
\begin{equation}\label{eq:heat_equation}
    \pder{x_i}\left(\sqrt{T}\pder[T]{x_i}\right) = 0.
\end{equation}
Here, it is taken into account that the thermal conductivity of a hard-sphere gas
is proportional to \(\sqrt{T}\).

A rigorous examination of the continuum limit shows that the infinitesimal thermal conduction term \(\Pder[q_j]{x_j}\)
can be of the same order as the thermal convective term \(\Pder[pv_j]{x_j}\) in the energy equation~\eqref{eq:energy}.
In such a case, when the Mach number is the same order of as the Knudsen number,
the heat-conduction equation~\eqref{eq:heat_equation} fails to describe
the correct temperature field. Moreover, additional thermal-stress terms of the second order of \(\Kn\)
appears in the momentum equation~\eqref{eq:momentum}.
These important modifications can be systematically considered only within the scope of kinetic theory.

\section{Asymptotic analysis of the Boltzmann equation}

A detailed mathematical derivation of the results introduced in this section
can be found in~\cite{Sone2002, Sone2007}.

Unless explicitly otherwise stated, all variables are further assumed to be dimensionless
by taking the following reference quantities:
the length \(L\), the pressure \(p^{(0)}\), the temperature \(T^{(0)}\),
the velocity \((2RT^{(0)})^{1/2}\), the heat-flow vector \(p^{(0)}(2RT^{(0)})^{1/2}\),
and the external force \(2RT^{(0)}/L\).
It is also convenient to deal with the modified Knudsen number
\[ k = \frac{\sqrt\pi}2\frac{\ell^{(0)}}{L} = \frac{m}{2\sqrt{2\pi} d_m^2 \rho^{(0)}L}. \]

\subsection{Fluid-dynamic-type equations}

The analysis presented below is based on the conventional Hilbert expansion
of the distribution function \(f\) and macroscopic variables \(h\)~\cite{Hilbert1912}:
\[ f = f_0 + f_1k + f_2k^2 + \cdots, \quad h = h_0 + h_1k + h_2k^2 + \cdots \]
under the additional assumption
\begin{equation}\label{eq:Mach_constraint}
    v_i = O(k)
\end{equation}
means that the Mach number is of the same order as the Knudsen number.
Moreover, let the external force be weak:
\begin{equation}\label{eq:Force_constraint}
    F_i = O(k^2).
\end{equation}
Conditions~\eqref{eq:Mach_constraint} and~\eqref{eq:Force_constraint} make the pressures
\(p_0\) and \(p_1\) constant due to the degenerated momentum equation~\eqref{eq:momentum}:
\begin{equation}
    \pder[p_0]{x_i} = 0, \quad \pder[p_1]{x_i} = 0.
\end{equation}

Then, the following set of equations for a time-independent case (\(\Pder{t} = 0\))
is obtained for variables \(T_0\), \(u_{i1} = p_0v_{i1}\), \(p_2^\dag\):
\begin{align}
    \pder{x_i}\left(\frac{u_{i1}}{T_0}\right) &= 0, \label{eq:asymptotic1} \\
    \pder{x_j}\left(\frac{u_{i1}u_{j1}}{T_0}\right)
        &-\frac{\gamma_1}2\pder{x_j}\left[\sqrt{T_0}\left(
            \pder[u_{i1}]{x_j} + \pder[u_{j1}]{x_i} - \frac23\pder[u_{k1}]{x_k}\delta_{ij}
        \right)\right] \notag\\
        &- \frac{\gamma_7}{T_0}\pder[T_0]{x_i}\pder[T_0]{x_j}\left(\frac{u_{j1}}{\gamma_2\sqrt{T_0}} - \frac{1}4\pder[T_0]{x_j}\right) \notag\\
        &= -\frac12\pder[p_2^\dag]{x_i} + \frac{p_0^2 F_{i2}}{T_0}, \label{eq:asymptotic2} \\
    \pder[u_{i1}]{x_i} &= \frac{\gamma_2}2\pder{x_i}\left(\sqrt{T_0}\pder[T_0]{x_i}\right), \label{eq:asymptotic3}
\end{align}
where
\begin{equation}\label{eq:dag_pressure}
    p_2^\dag = p_0 p_2
        + \frac{2\gamma_3}{3}\pder{x_k}\left(T_0\pder[T_0]{x_k}\right)
        - \frac{\gamma_7}{6}\left(\pder[T_0]{x_k}\right)^2.
\end{equation}
The fluid-dynamic-type equations \eqref{eq:asymptotic1}--\eqref{eq:asymptotic3}
are comparable to the Navier--Stokes equations for a compressible gas (\(\rho_0 = p_0/T_0\)).
According to the pioneering authors~\cite{Kogan1971, Kogan1976, ExperimentsNTFS2003},
they are called \emph{slow nonisothermal (SNIT) flow equations}.
The formal difference consists in the additional thermal-stress term.
Comparing it with \(p_0^2F_{i2}/T_0\), we find that
\begin{equation}\label{eq:force}
    k^2\frac{\gamma_7}{p_0^2}\pder[T_0]{x_i}\pder[T_0]{x_j}\left(\frac{u_{j1}}{\gamma_2\sqrt{T_0}} - \frac{1}4\pder[T_0]{x_j}\right)
\end{equation}
is the force acting on unit mass of the gas.
For a hard-sphere gas at rest, this force is opposite to the temperature gradient direction.

It should be noted that \(p_2^\dag\) is not included in the equation of state and
therefore is determined up to a constant factor.
Term \(\Pder[p_2^\dag]{x_i}\) participates in the system as the pressure
in the Navier--Stokes set for an incompressible gas.
This fact specifies the class of suitable numerical methods
for solving \eqref{eq:asymptotic1}--\eqref{eq:asymptotic3}.

For a hard-sphere gas, the transport coefficients are
\begin{alignat*}{2}
    \gamma_1 &= 1.270042427, &\quad \gamma_2 &= 1.922284066, \\
    \gamma_3 &= 1.947906335, &\quad \gamma_7 &= 1.758705.
\end{alignat*}
The first two are connected to the dimensional transport coefficients in the following way:
\begin{equation}
    \mu = \gamma_1\sqrt{T_0} \frac{p^{(0)}L}{\sqrt{2RT^{(0)}}} k, \quad
    \lambda = \frac{5\gamma_2}{2}\sqrt{T_0} \frac{p^{(0)}RL}{\sqrt{2RT^{(0)}}} k,
\end{equation}
and \(\gamma_7\) is related to the \emph{nonlinear thermal-stress flow}.
Other molecular models are easily applied by replacing these coefficients.
Appropriate formulas for their numerical evaluation can be found in~\cite{SoneBobylev96, Sone2002, Sone2007}.

\subsection{Boundary conditions}

The fluid-dynamic-type boundary conditions based on the diffuse-reflection boundary conditions
for Boltzmann equation look as follows:
\begin{gather}
    T_0 = T_{B0}, \label{eq:bound:T} \\
    \left\{
    \begin{aligned}
        & \frac{(u_{j1}-u_{Bj1})}{\sqrt{T_{B0}}}(\delta_{ij}-n_in_j) =
            -K_1\pder[T_{B0}]{x_j}(\delta_{ij}-n_in_j), \\
        & u_{j1}n_j = 0.
    \end{aligned}
    \right. \label{eq:bound:v}
\end{gather}
Here, \(n_i\) is the unit normal vector to the boundary, pointing into the gas.
\(u_{Bj1}k\) is the velocity of the boundary, and \(T_{B0}\) is its temperature.
\(K_1\) is the slip coefficient related to the \emph{thermal creep flow}.
For a hard-sphere gas, \[ K_1 = -0.6463. \]
Thus, the direction of the thermal creep flow coincides with the temperature gradient.

The solution of the boundary-value problem cannot be expressed in terms of the fluid-dynamic-type
solution due to the sharp variation in the distribution function near the boundary.
Therefore, it is necessary to introduce the so-called Knudsen-layer correction:
\begin{equation}
    f = f_{FD} + f_K, \quad h = h_{FD} + h_K.
\end{equation}
The fluid-dynamic part \(f_{FD}\) is the solution presented above,
and \(f_K\) decays exponentially with the distance to the boundary \(f_K = O\left(e^{-\eta}\right)\).
The relation \( x_i = k\eta n_i(s_1,s_2) + x_{Bi}(s_1, s_2) \) gives a coordinate transformation
from \((x_1,x_2,x_3)\) to \((\eta,s_1,s_2)\).

\begin{figure}[ht]
    \centering
    \begin{minipage}{.48\textwidth}
        \centering
        \includegraphics{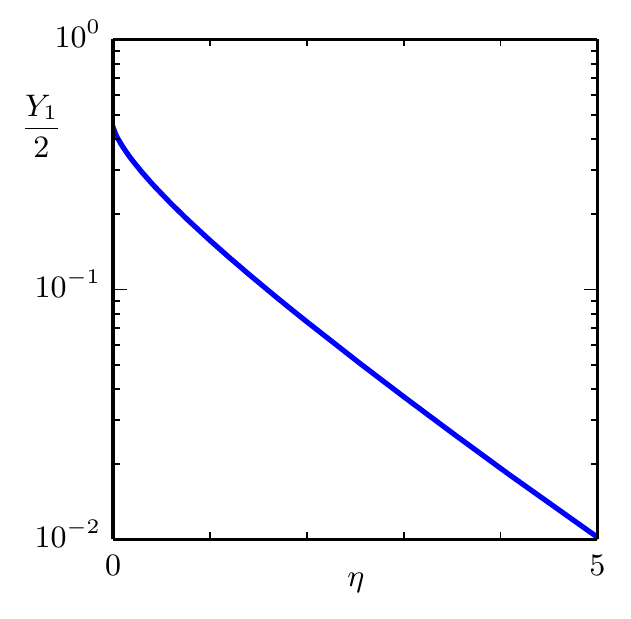}
        \caption{The function of the Knudsen layer \(Y_1(\eta)/2\) for a hard-sphere gas.}
        \label{fig:Y1}
    \end{minipage}
    \quad
    \begin{minipage}{.48\textwidth}
        \centering
        \includegraphics{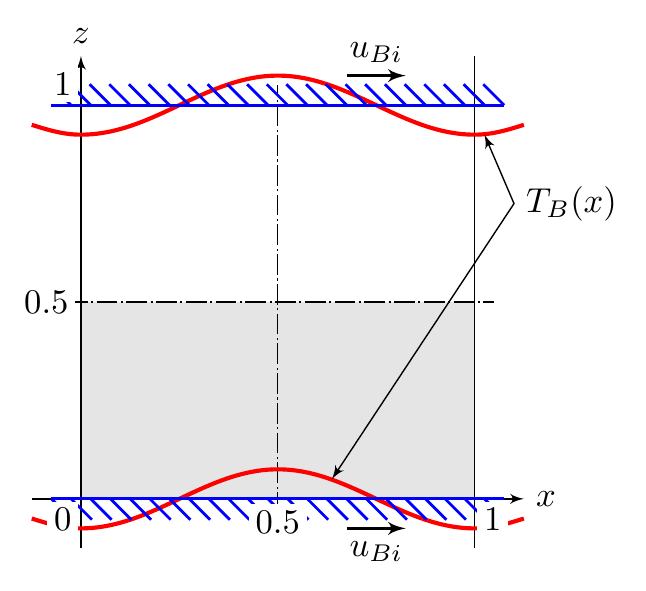}
        \vspace{13pt}
        \caption{Geometry of the problem: the gas is placed between two parallel plates
            with a periodic temperature distribution \(T_B(x)\) and a velocity \(u_{Bi}\).}
        \label{fig:geometry}
    \end{minipage}
\end{figure}

Due to the infinitesimal Mach number~\eqref{eq:Mach_constraint}, \(f_K\) is expanded
in a power series of \(k\), starting from the first order:
\[ f_K = f_{K1} k + f_{K2} k ^ 2 + \cdots \]
The Knudsen layer introduces a correction to \(u_{i1}\) for a hard-sphere gas:
\begin{equation}
    \left\{
    \begin{aligned}
        & \frac{u_{jK1}}{\sqrt{T_{B0}}}(\delta_{ij}-n_in_j) =
            -\frac12\pder[T_{B0}]{x_j} Y_1\left(\frac\eta{T_{B0}}\right) (\delta_{ij}-n_in_j), \\
        & u_{jK1}n_j = 0.
    \end{aligned}
    \right. \label{eq:bound:v_K}
\end{equation}
The function \(Y_1(\eta)\) is tabulated, for example, in~\cite{Sone2002, Sone2007} and is shown in Fig.~\ref{fig:Y1}.

\subsection{Force acting on a closed body}

The stress tensor \(p_{ij}\) and the heat-flow vector \(q_i\) in a gas are obtained as follows:
\begin{alignat}{2}
    p_{ij0} &= p_0\delta_{ij}, && \quad q_{i0} = 0, \label{eq:stress0}\\
    p_{ij1} &= p_1\delta_{ij}, && \quad q_{i1} = -\frac{5\gamma_2}4\sqrt{T_0}\pder[T_0]{x_i}, \label{eq:stress1}\\
    p_{ij2} &= p_2\delta_{ij}
                &&- \frac{\gamma_1}{p_0}\sqrt{T_0}\left(
                \pder[u_{i1}]{x_j} + \pder[u_{j1}]{x_i} - \frac23\pder[u_{k1}]{x_k}\delta_{ij}
            \right) \notag\\
        & &&+ \frac{\gamma_3}{p_0} T_0\left(
                \pderder[T_0]{x_i}{x_j} - \frac13\pderdual[T_0]{x_k}\delta_{ij}
            \right) \notag\\
        & &&+ \frac{\gamma_3 - \gamma_7}{p_0}\left[
                \pder[T_0]{x_i}\pder[T_0]{x_j} - \frac13\left(\pder[T_0]{x_k}\right)^2\delta_{ij}
            \right]. \label{eq:stress2}
\end{alignat}
The formula for \(q_{i1}\) coincides with Fourier's law~\eqref{eq:Fourier_law},
but \(p_{ij2}\) contains the thermal-stress terms in addition to the viscous one.
Due to the nonuniform stresses, an internal force of the second order of \(k\) per unit area,
\(F_{i2} = -p_{ij2}n_j\), acting on a body, arises in the gas.
The unit normal vector \(n_i\) points into the gas as before.

With the help of divergence theorem, the second-order differential term in Eq.~\eqref{eq:stress2}
can be transformed during the integration over the body surface in the following way:
\begin{align*}
    \oint_S T_0\pderder[T_0]{x_i}{x_j} n_j\dd{S}
        &= \oint_S \pder{x_i} \left( T_0\pderder[T_0]{x_i}{x_j} \right) n_j\dd{S}
        - \oint_S \pder[T_0]{x_i}\pder[T_0]{x_j} n_j\dd{S} \\
        &= \int_V \pder{x_i}\pder{x_j} \left( T_0 \pder[T_0]{x_j} \right) \dd{V}
        - \oint_S \pder[T_0]{x_i}\pder[T_0]{x_j} n_j\dd{S} \\
        &= \oint_S \left(\pder[T_0]{x_j}\right)^2 n_i\dd{S}
        + \oint_S T_0\pderdual[T_0]{x_j} n_i\dd{S}
        - \oint_S \pder[T_0]{x_i}\pder[T_0]{x_j} n_j\dd{S},
\end{align*}
where the integration is carrying out over the body volume \(V\) and the body surface \(S\).
A part of the viscosity term can also be modified in the same manner:
\begin{equation*}
    \oint_S \sqrt{T_0}\pder[u_{j1}]{x_i} n_j\dd{S}
        = \oint_S \pder{x_i} \left( \sqrt{T_0} u_{j1} \right) n_j\dd{S}
        = \int_V \pder{x_i}\pder{x_j} \left( \sqrt{T_0} u_{j1} \right) \dd{V}
        = \frac{3}{2}\oint_S \sqrt{T_0} \pder[u_{j1}]{x_j} n_i\dd{S}.
\end{equation*}
Here, the boundary condition~\eqref{eq:bound:v} and the continuity equation~\eqref{eq:asymptotic1} are used.

Owing to the above formulas, the total force acting on a closed body
is expressed in terms of \(T_0\), \(u_{i1}\), \(p^\dag_2\) in the following form:
\begin{align}\label{eq:force:terms_general}
    p_0 \oint_S F_{i2} \dd{S} =
        &- \oint_S p_2^\dag n_i \dd{S} \notag\\
        &+ \frac{5}{6}\gamma_1 \oint_S \sqrt{T_{B0}} \pder[u_{j1}]{x_j} n_i \dd{S}
        + \gamma_1 \oint_S \sqrt{T_{B0}} \pder[u_{i1}]{x_j} n_j \dd{S} \notag\\
        &- \frac{\gamma_7}{2} \oint_S \left(\pder[T_0]{x_j}\right)^2 n_i \dd{S}
        + \gamma_7 \oint_S \pder[T_0]{x_i}\pder[T_0]{x_j} n_j \dd{S}.
\end{align}
In particular, a body at rest (\(u_{Bi} =0 \)) with a uniform boundary temperature (\(T_B = \mathrm{const}\))
is acted on by the force consisting on the following components:
\begin{equation}\label{eq:force:terms}
    p_0 \oint_S F_{i2} \dd{S} =
        - \oint_S p_2^\dag n_i \dd{S} \\
        + \gamma_1 \sqrt{T_{B0}} \oint_S \pder[u_{i1}]{x_j} n_j \dd{S}
        + \frac{\gamma_7}{2} \oint_S \left(\pder[T_0]{x_j}\right)^2 n_i \dd{S}.
\end{equation}
The Knudsen-layer correction is excluded from the consideration since have a zero contribution to the total force.
This fact can be easily proved by moving the surface of integration outside the Knudsen layer.

\subsection{Some remarks on the continuum limit}

As one can see from~\eqref{eq:asymptotic3}, a gas is not always described correctly
by the heat-conduction equation~\eqref{eq:heat_equation}.
If \(u_{i1} = 0\), Eq.~\eqref{eq:asymptotic3} converges to Eq.~\eqref{eq:heat_equation}.
In the absence of an external force, there are several reasons when this condition can be violated:
\begin{enumerate}
    \item The boundary is moving: \(u_{B1i} \neq 0 \).
    \item The boundary temperature is not uniform: \(\Pder[T_{B0}]{x_i} \neq 0 \).
    \item The isothermal surfaces are not parallel:
        \begin{equation}\label{eq:equilibrium}
            e_{ijk}\pder[T_0]{x_j}\pder{x_k}\left(\pder[T_0]{x_l}\right)^2 \neq 0.
        \end{equation}
\end{enumerate}
The first two cases are directly determined by the boundary conditions,
and the third one has only an implicit impact from them.

The set of equations~\eqref{eq:asymptotic1}--\eqref{eq:asymptotic3} revealed the following interesting fact.
In the continuum limit, the velocity field tends to zero, proportional to the Knudsen number,
due to~\eqref{eq:Mach_constraint}, however infinitesimally weak flows
have a finite impact on the temperature field.
This effect arises in the continuum world owing to the additional nonlinear thermal-stress term
in the momentum equation~\eqref{eq:asymptotic2}.
Some authors refer to it as a \emph{ghost effect}~\cite{Sone2002, Sone2007}.

\section{Numerical simulation and discussions}

\begin{figure}[ht]
    \centering
    \includegraphics{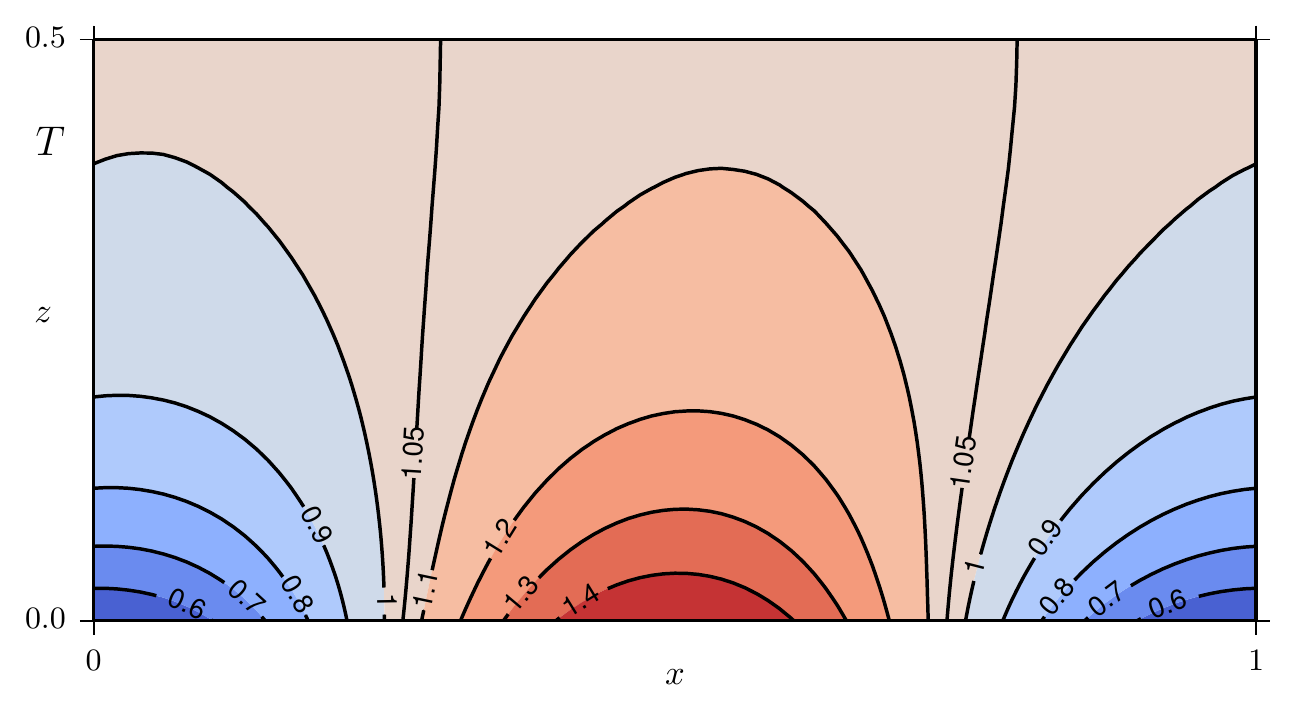}
    \caption{The temperature field \(T_0\) obtained from the SNIT flow equations.
        The isothermal lines are placed above the contour lines.}
    \label{fig:moving:T_asym}
\end{figure}

\begin{figure}[ht]
    \centering
    \includegraphics{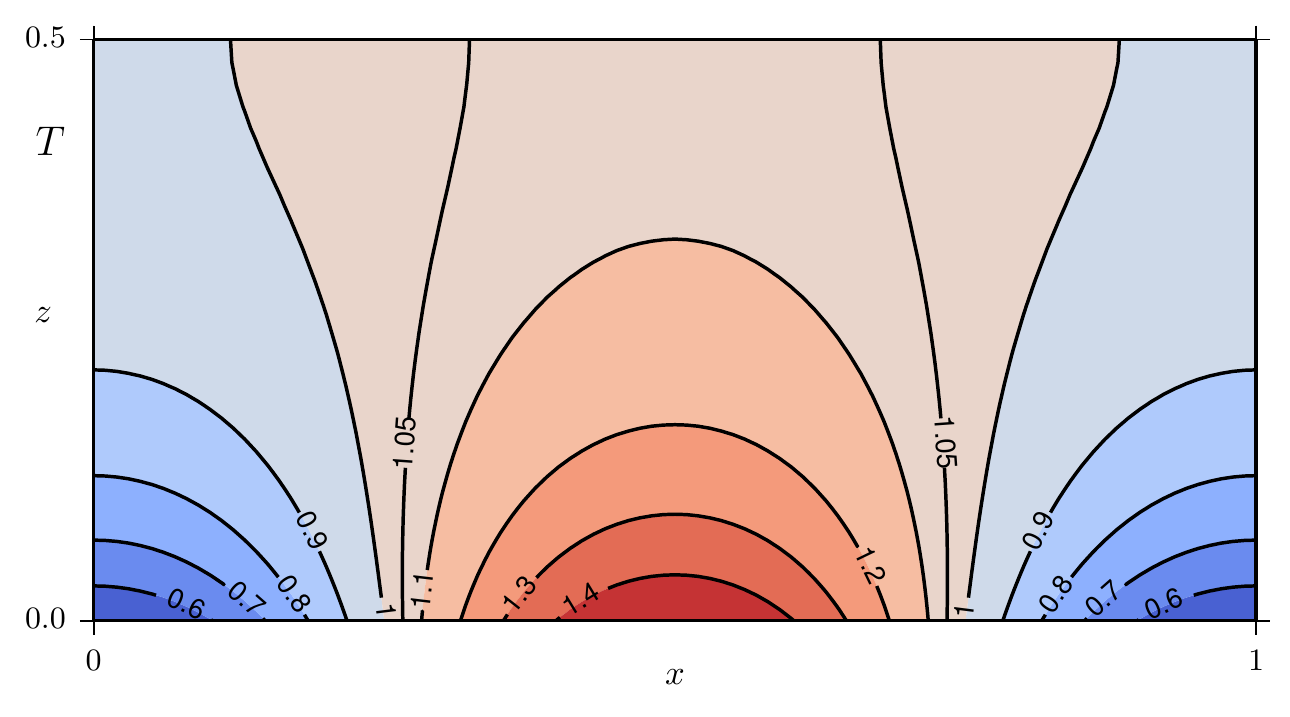}
    \caption{Temperature field \(T\) obtained from the heat-conduction equation.
        The isothermal lines are placed above the contour lines.}
    \label{fig:moving:T_heat}
\end{figure}

The present results were obtained using the special-purpose solver of
the SNIT flow equations~\eqref{eq:asymptotic1}--\eqref{eq:asymptotic3}.
It has been developed on the open-source CFD platform OpenFOAM\textregistered{}~\cite{OpenFOAM1998},
based on the finite-volume approach for evaluating partial derivatives.
The continuity equation~\eqref{eq:asymptotic1}, along with the momentum equation~\eqref{eq:asymptotic2},
is solved using the implicit conservative SIMPLE algorithm~\cite{SIMPLE}.
A detailed description of the similar scheme for gas mixtures can be found in~\cite{Laneryd2007}.
According to the naming convention adopted in OpenFOAM\textregistered{},
the introduced solver can be called as \verb+snitSimpleFoam+.

Spatial grids are generated with the open-source package GMSH~\cite{GMSH}.
They are nonuniform and condensed in areas of a large temperature variation.
From \(10^3\) to \(10^6\) cells are used to achieve the residual of the solution
that is not greater than \(10^{-6}\) for all considered problems.
The temperature fields are evaluated at least with precision \(10^{-9}\).
A modern personal computer (3.40GHz CPU) has been used to simulate presented problems.
For the most detailed spatial grids, it takes about several minutes to reach the solution.
This fact provides a great performance for parametric studies.

OpenFOAM\textregistered{} uses Kitware's ParaView\textregistered{} as a primary visualization system of results,
but for the present paper another open-source tool, Matplotlib, has been chosen to create color vector illustrations.

All transport coefficients are taken in compliance with the hard-sphere model.
Complete diffuse reflection is assumed from a solid boundary.
The external force is taken to be zero.

\subsection{Gas between two parallel plates}

Consider a plane periodic geometry shown in Fig.~\ref{fig:geometry}.
A gas is placed between two infinite parallel plates,
both with the periodic distribution of the temperature and the small velocity:
\begin{equation}
    T_B = 1-\alpha\cos(2\pi x), \quad u_{Bi} = (\beta\Kn,0,0).
\end{equation}
Due to the symmetry, the computational domain represents a rectangle \(0<x<1, 0<z<1/2\)
(a gray background in Fig.~\ref{fig:geometry}).
The problem is simulated for
\[ \alpha=1/2, \quad \beta = 1 \]
to compare with the reference solution in~\cite{SoneBobylev96}.
Complete agreement between these results is a good verification of the solver validity.

\begin{figure}
    \centering
    \includegraphics{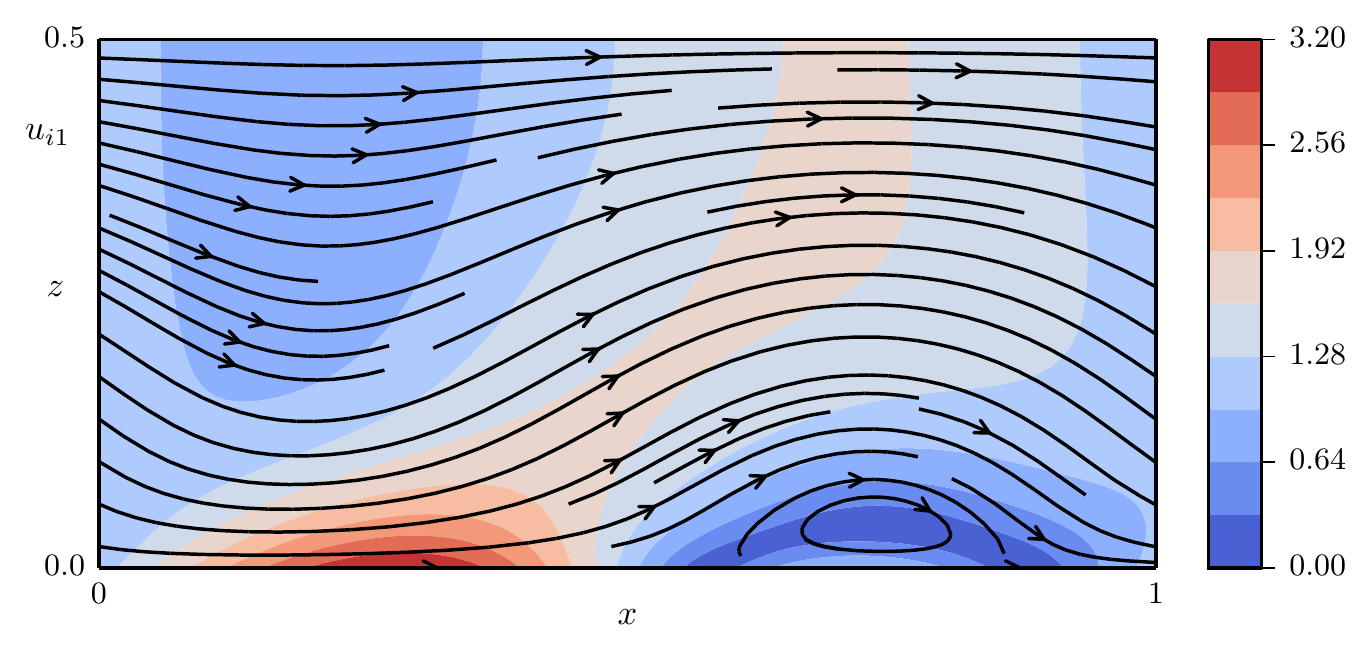}
    \caption{The fluid-dynamic part of the velocity field \(u_{FDi1}\).
        Curves with arrows indicate the direction, contour lines show the magnitude.}
    \label{fig:moving:fluid}
\end{figure}

\begin{figure}
    \centering
    \includegraphics{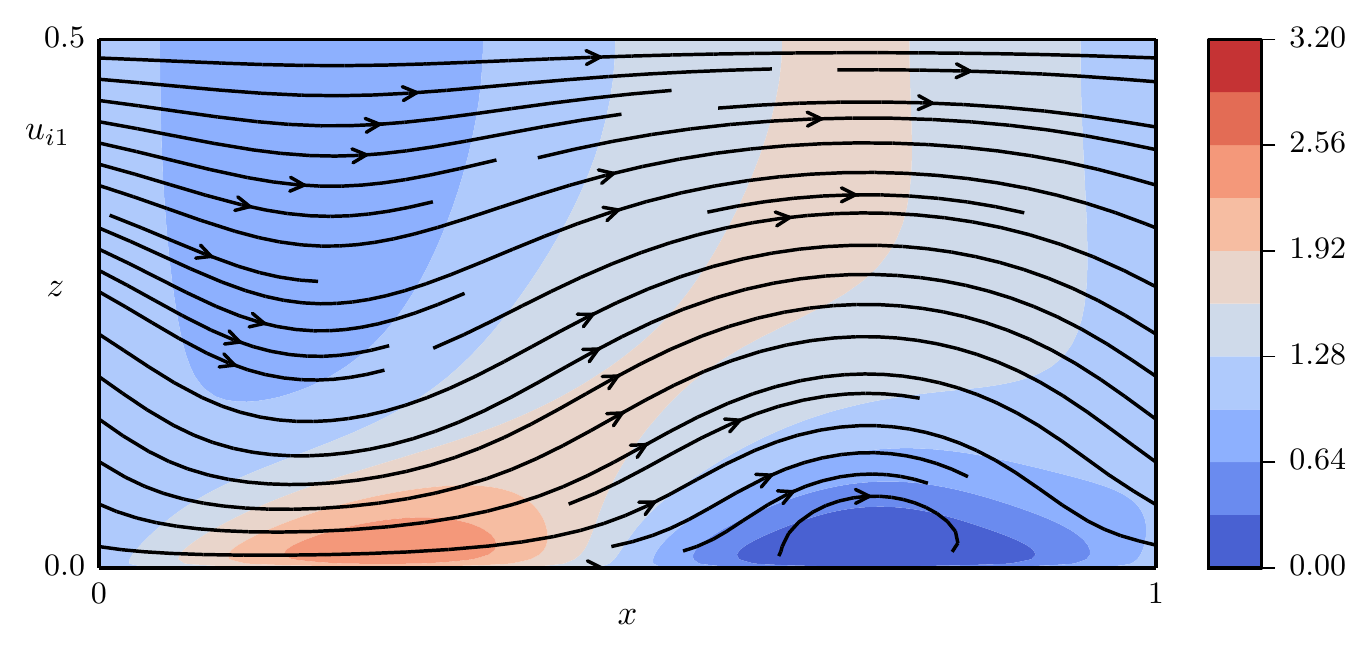}
    \caption{The velocity field \(u_{i1}\) with the Knudsen-layer correction for \(\Kn=0.01\).
        Curves with arrows indicate the direction, contour lines show the magnitude.}
    \label{fig:moving:kn001}
\end{figure}

The temperature field (Fig.~\ref{fig:moving:T_asym}) is shown
in comparison with the solution of the heat-conduction equation (Fig.~\ref{fig:moving:T_heat}).
In the continuum limit, \(T=T_0\), and one can see that the classical fluid-dynamic solution
has a finite distinction from the kinetic one.

The velocity field \(u_{i1}\) is shown in Fig.~\ref{fig:moving:fluid} and Fig.~\ref{fig:moving:kn001}.
The Knudsen-layer correction~\eqref{eq:bound:v_K} is taken into account in the latter figure.
For small \(k\), a gas flow along the boundary temperature gradient occurs
in the thin Knudsen layer. This flow, which is of the first order of \(k\), is called
the \emph{thermal creep flow}.

\subsection{Gas between two cylinders and spheres}

\begin{figure}
    \centering
    \includegraphics{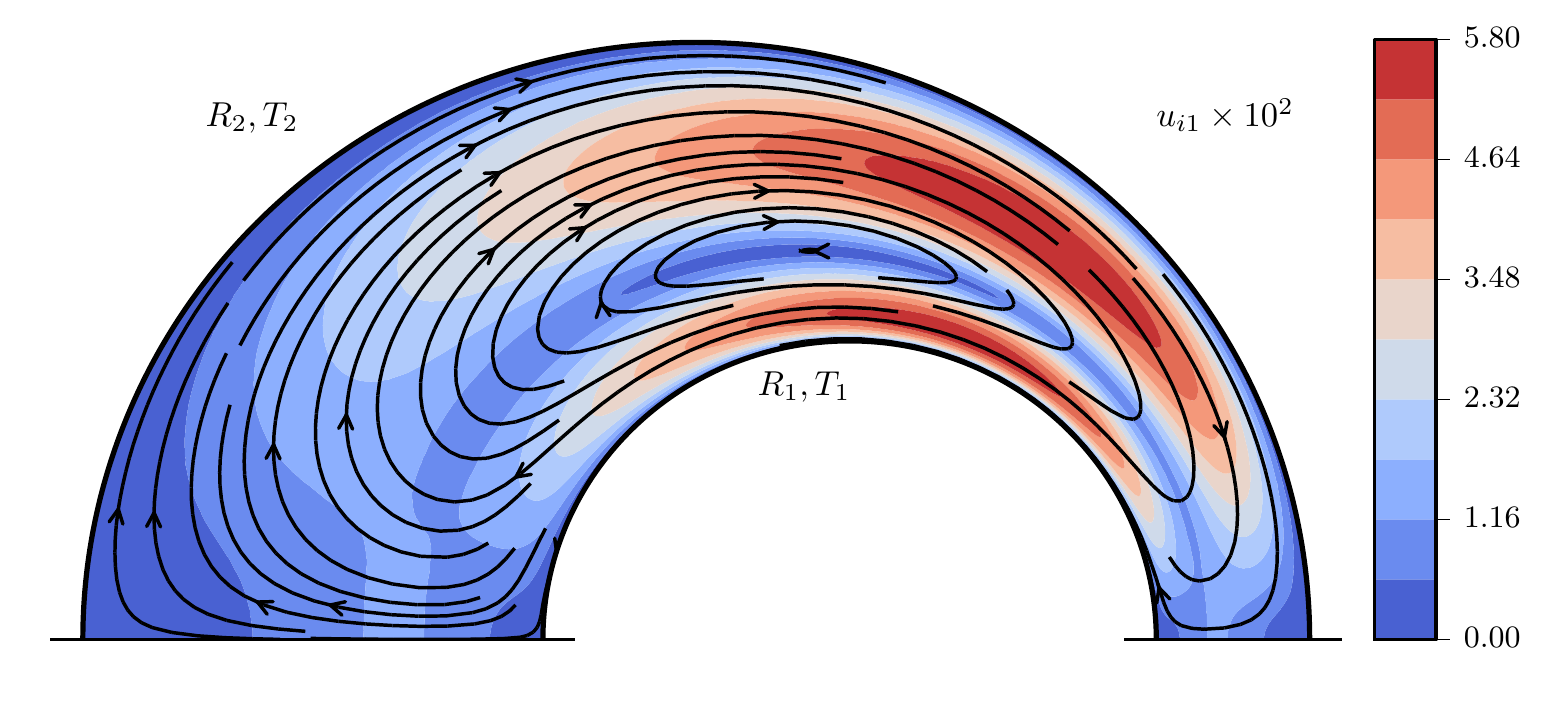}
    \caption{The velocity field \(u_{i1}\) between two noncoaxial cylinders for \(\tau=4\).
        Curves with arrows indicate the direction, contour lines show the magnitude.}
    \label{fig:cylinders}
\end{figure}

\begin{figure}
    \centering
    \includegraphics{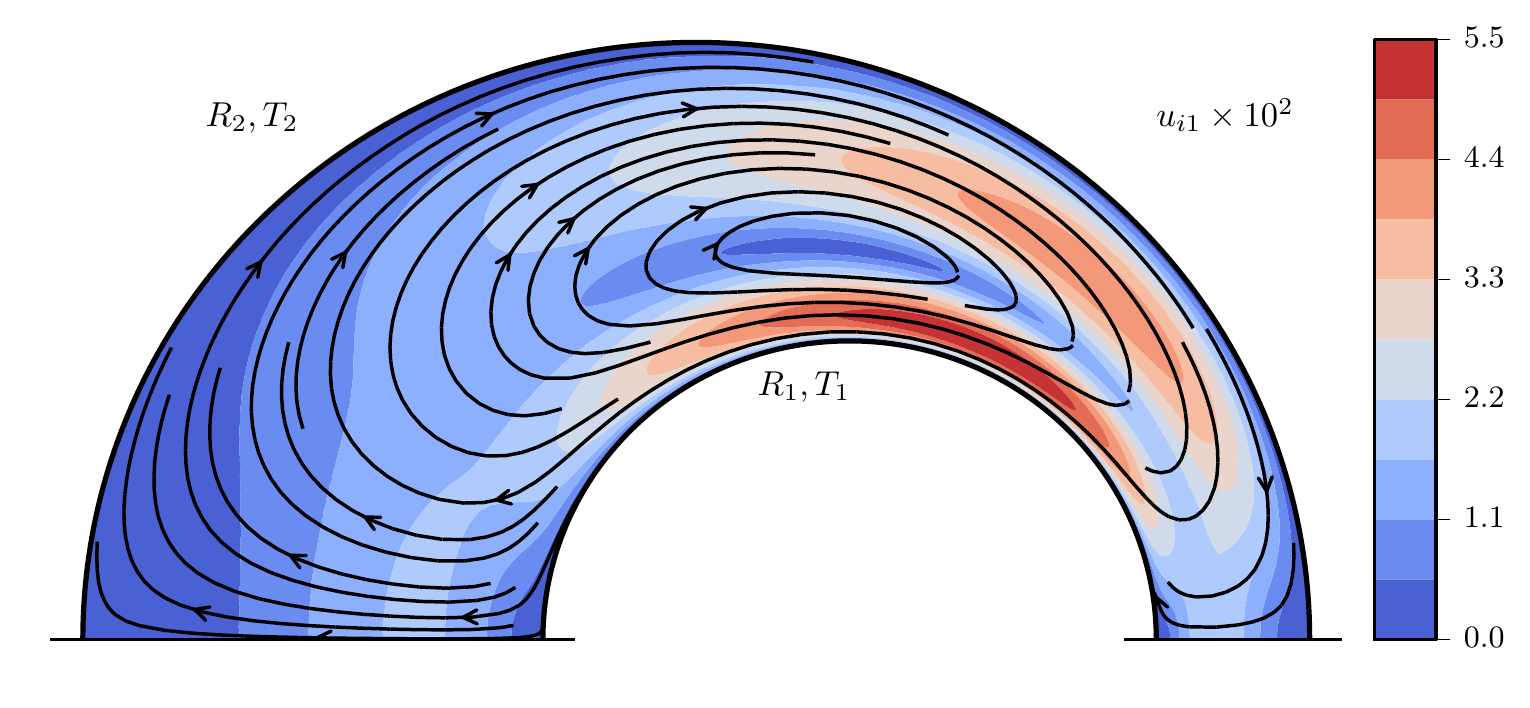}
    \caption{The velocity field \(u_{i1}\) between two nonconcentric spheres for \(\tau=4\).
        Curves with arrows indicate the direction, contour lines show the magnitude.}
    \label{fig:spheres}
\end{figure}

Now, consider the case where there is no temperature gradient on the surface of the surrounding bodies at rest.
The Navier--Stokes equations with any slip boundary conditions have a trivial solution \(u_i = 0\),
which is, however, not valid for~\eqref{eq:asymptotic1}--\eqref{eq:asymptotic3}.
As stated above, even under such boundary conditions, the \emph{nonlinear thermal-stress flow}
can emerge due to nonparallelism of the isothermal surfaces~\eqref{eq:equilibrium}.

Consider two cylinders of radii \(R_1 = 1\) and \(R_2 = r\)
with temperatures \(T_1 = 1\) and \(T_2 = 1+\tau\) respectively.
The cylinder axes are parallel; the distance between them is equal to \(d\) along \(x\) axis.

Fig.~\ref{fig:cylinders} presents the results of the numerical simulation for
\[ r = 2, \quad d = 0.5, \quad \tau = 4. \]
A gas between nonconcentric spheres is also simulated in the same geometry (Fig.~\ref{fig:spheres}).
In comparison with the previous example, the velocity field \(u_i\) is two orders of magnitude weaker.
This is the primary reason why an experimental study of nonlinear thermal-stress flow is so complicated.

\begin{figure}[ht]
    \centering
    \begin{minipage}{.48\textwidth}
        \centering
        \includegraphics{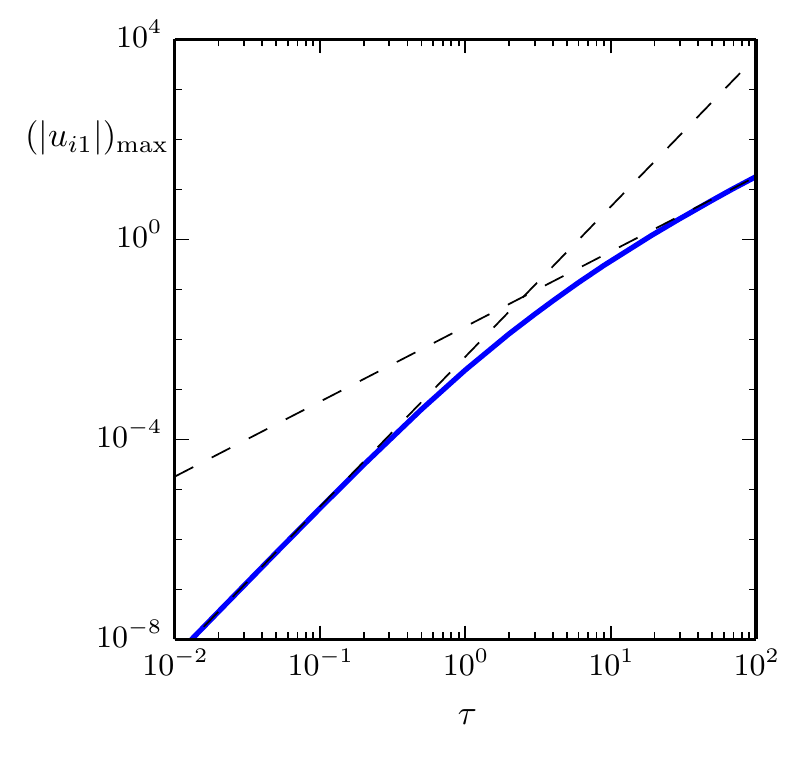}
        \caption{The maximum magnitude of \(u_{i1}\) versus \(\tau\) for \(d=0.5\).
                When \(\tau\) goes to zero, it is proportional to \(\tau^3\).
                For large \(\tau\), it is proportional to \(\tau^{3/2}\).}
        \label{fig:tau:maxU}
    \end{minipage}
    \quad
    \begin{minipage}{.48\textwidth}
        \centering
        \includegraphics{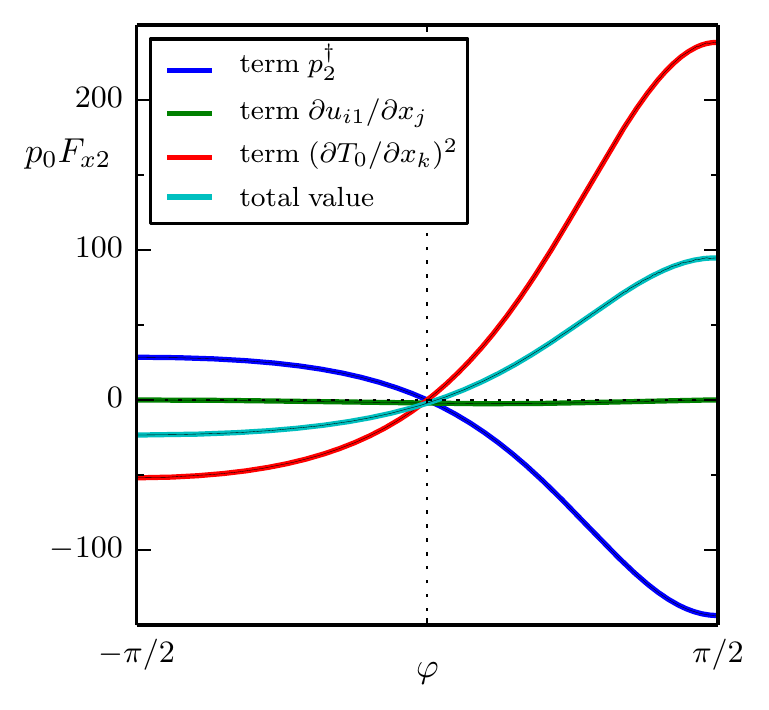}
        \caption{The profile of the components of the acting force \(F_{x2}\) along
                the inner circle in polar coordinates for \(d=0.5\) and \(\tau=4\).
                \(\varphi = -\pi/2\) corresponds to the point~\(x=d-1\),
                \(\varphi = \pi/2\) corresponds to the point~\(x=d+1\).}
        \label{fig:terms:inner}
    \end{minipage}
\end{figure}

The nonlinear thermal-stress flow decreases as a cubic function of small temperature variations
owing to the fact that the temperature gradient is included in the thermal-stress force
as a cubic term [see Eq.~\eqref{eq:force}].
Therefore, when square of temperature variations becomes comparable to the Knudsen number,
one should takes into account the gas effects of the second order of Knudsen number~\cite{Sone1989Noncoaxial}.
In the presence of large temperature variations,
the slope of the relation becomes less steep since the viscosity of a hard-sphere gas increases with temperature.
This dependence is demonstrated in Fig.~\ref{fig:tau:maxU}.

Let us now examine the force acting on the cylinders.
As noted above, \(p_{ij2}\) is determined up to the constant factor
from the system~\eqref{eq:asymptotic1}-\eqref{eq:asymptotic3}.
This constant factor makes no contribution to the total force acting on a body,
but to evaluate a force per unit area, we have to impose an additional condition
on \(p_{ij2}\), for example, the following one:
\begin{equation}\label{eq:dag_condition}
    \int_\Omega p^\dag_{ij2}\dd{V} = 0,
\end{equation}
where the three-dimensional integration is carried out over the whole domain \(\Omega\) where the gas presents.

\begin{figure}[ht]
    \centering
    \begin{minipage}{.48\textwidth}
        \centering
        \includegraphics{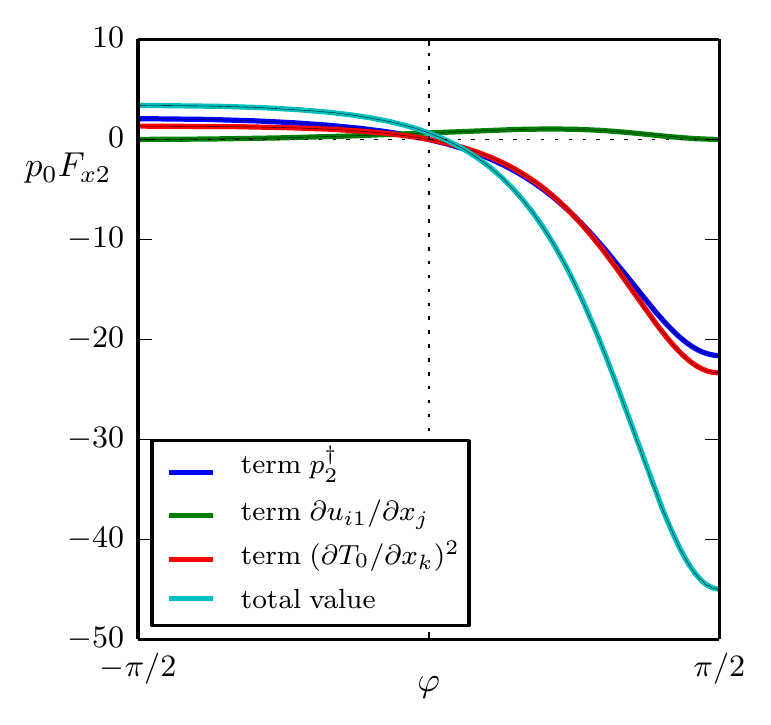}
        \caption{The profile of the components of the acting force \(F_{x2}\) along
                the outer circle in polar coordinates for \(d=0.5\) and \(\tau=4\).
                \(\varphi = -\pi/2\) corresponds to the point~\(x=-r\),
                \(\varphi = \pi/2\) corresponds to the point~\(x=r\).}
        \label{fig:terms:outer}
    \end{minipage}
    \quad
    \begin{minipage}{.48\textwidth}
        \centering
        \includegraphics{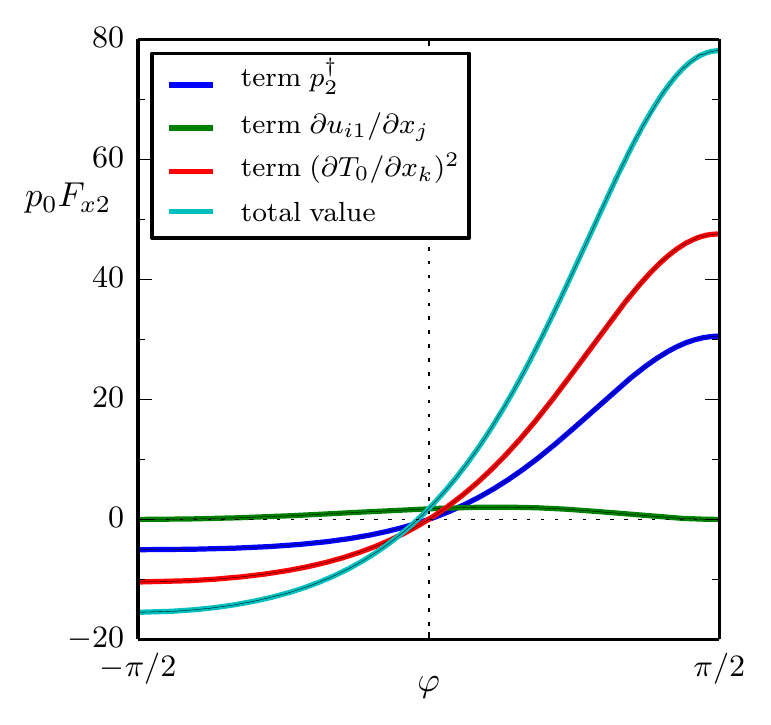}
        \caption{The profile of the components of the acting force \(F_{x2}\)
                along the inner circle in polar coordinates
                for \(d=0.5\) and the swapped temperatures \(T_1 = 5\), \(T_2 = 1\).
                \(\varphi = -\pi/2\) corresponds to the point~\(x=d-1\),
                \(\varphi = \pi/2\) corresponds to the point~\(x=d+1\).}
        \label{fig:terms:inner-swap}
    \end{minipage}
\end{figure}

The contribution of each term of~\eqref{eq:force:terms} to the total value is showed
in Fig.~\ref{fig:terms:inner},~\ref{fig:terms:outer},~\ref{fig:terms:inner-swap}.
Note that the depicted profile of the total value does not correspond to the real profile of the force \(F_{i2}\)
acting on the unit area, but only to the sum of all terms in~\eqref{eq:force:terms}.
Actually, an evaluation of the acting force at a certain point requires the calculation
of the mixed second-order partial derivatives on the boundary,
which is a complicated problem within the finite-volume method.

\begin{figure}[ht]
    \centering
    \begin{minipage}{.48\textwidth}
        \centering
        \includegraphics{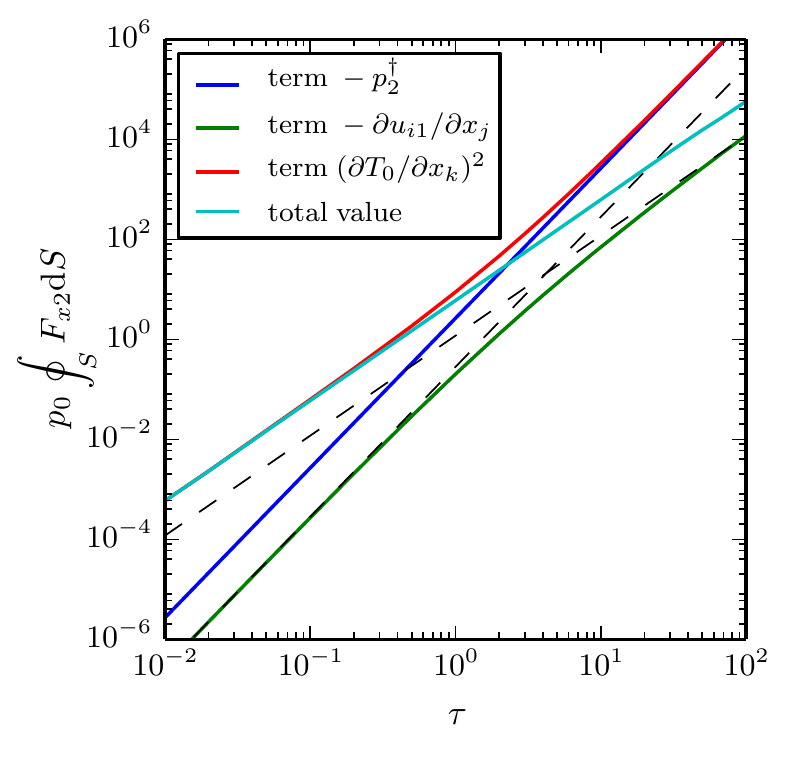}
        \caption{The total force acting on the inner cylinder versus \(\tau\) for \(d=0.5\).}
        \label{fig:tau:force-inner}
    \end{minipage}
    \quad
    \begin{minipage}{.48\textwidth}
        \centering
        \includegraphics{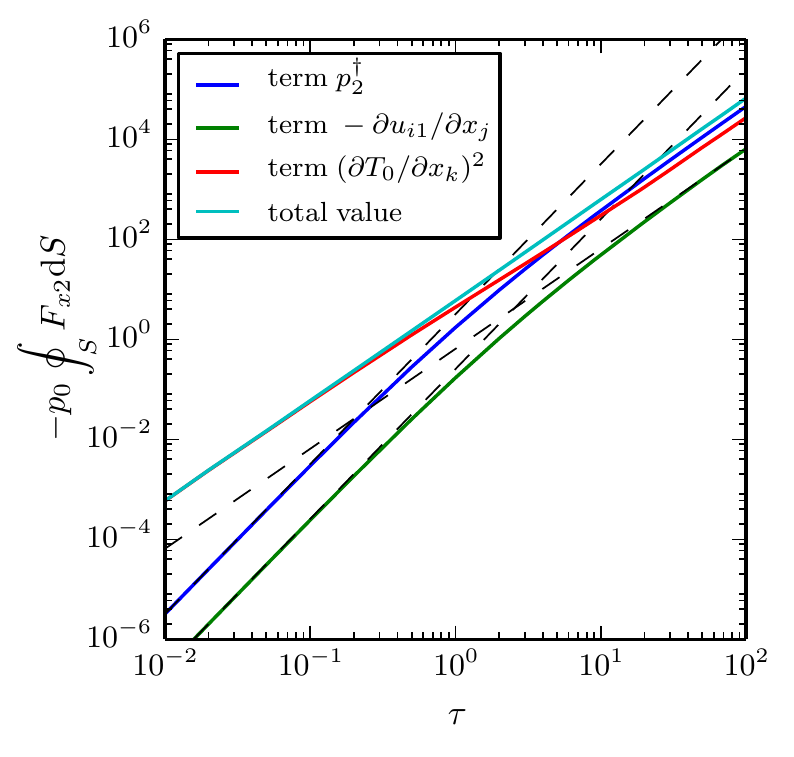}
        \caption{The total force acting on the outer cylinder versus \(\tau\) for \(d=0.5\).}
        \label{fig:tau:force-outer}
    \end{minipage}
\end{figure}

The cylinders are attracted with a force proportional to \(\tau^2\) over the entire temperature range,
although individual components do not obey this relation (Fig.~\ref{fig:tau:force-inner}, \ref{fig:tau:force-outer}).
For a hard-sphere gas, temperature gradient on the inner cylinder increases proportional to \(\tau^{3/2}\),
but it is compensated by the hydrostatic pressure \(p^\dag_2\).
So the pressure term is opposite to the thermal-stress one on the inner cylinder
(Fig.~\ref{fig:terms:inner}, \ref{fig:tau:force-inner}).
The viscous term makes a minor contribution to the total force.
If we swap the temperatures \(T_1\) and \(T_2\) (Fig.~\ref{fig:terms:inner-swap}),
the total force changes neither its direction nor its magnitude.

\begin{figure}[ht]
    \centering
    \begin{minipage}{.48\textwidth}
        \centering
        \includegraphics{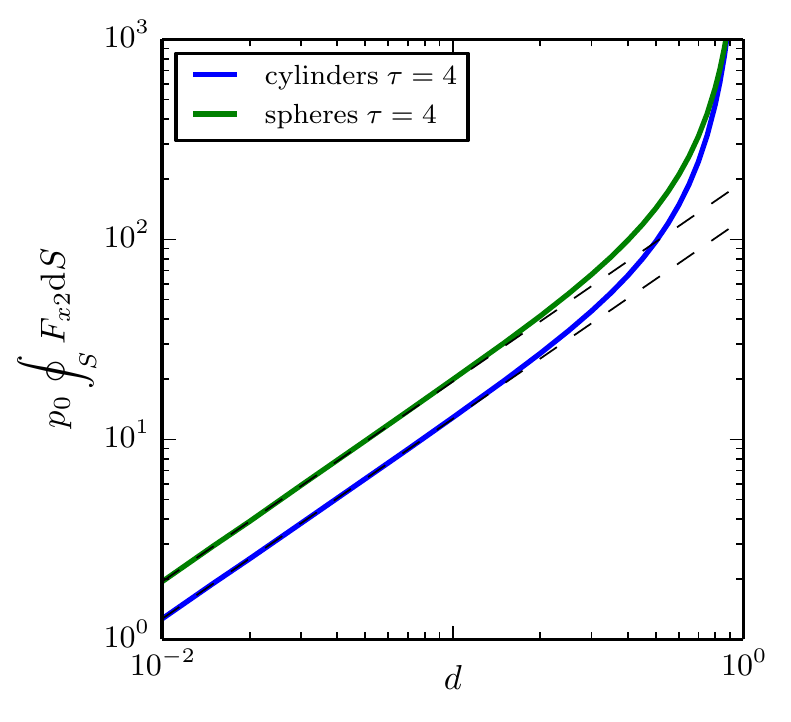}
        \caption{The total force acting on the inner cylinder (sphere)
                versus the distance between their axes~\(d\).
                The dashed lines correspond to a linear relation.}
        \label{fig:forces:total}
    \end{minipage}
    \quad
    \begin{minipage}{.48\textwidth}
        \centering
        \includegraphics{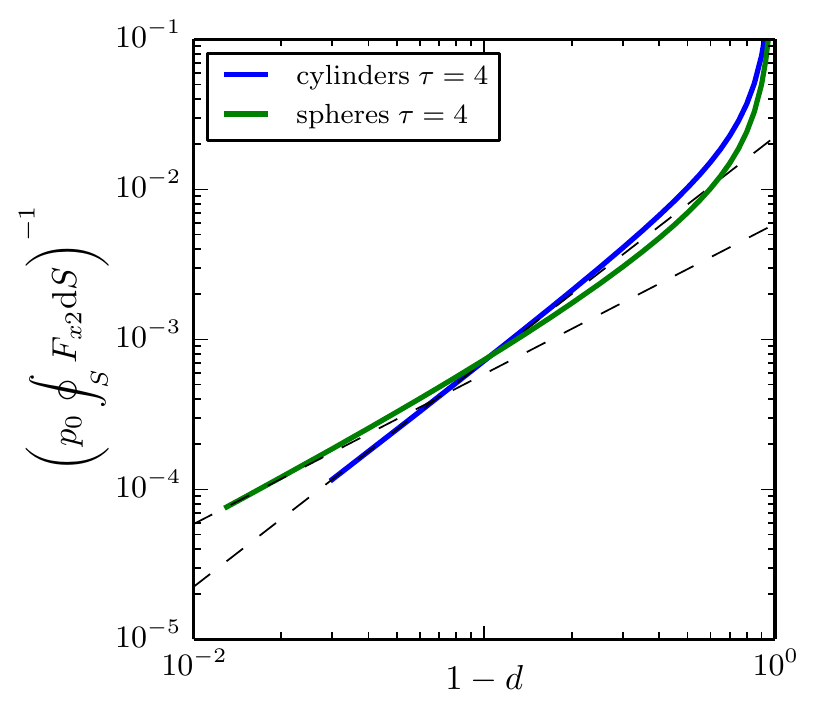}
        \caption{The total force acting on the inner cylinder (sphere)
                versus the distance between their axes~\(d\).
                The dashed lines correspond to a linear relation for spheres
                and to a power of \(3/2\) for cylinders.}
        \label{fig:forces:inverse}
    \end{minipage}
\end{figure}

The acting force is similar to the electrostatic force with potential \(T\),
so the attractive force can be considered as in a cylindrical (spherical) capacitor.
One can write down the total force in the following form:
\begin{equation}
    p_0\oint_S F_{x2}\dd{S} = \pder[C]{d} \tau^2,
\end{equation}
where \(C\) is an analog of the capacitance depending on the distance between the axes \(d\).
The electrostatic solution gives the following formulas for cylinders and spheres~\cite{Smythe1968Electricity}:
\begin{equation}
    C_\mathrm{cyl} \propto \frac1{\theta}, \quad
    C_\mathrm{sph} \propto  \sum_{n=1}^\infty \frac{R_1 R_2 \sinh\theta} {R_2\sinh n\theta - R_1\sinh (n-1)\theta}, \quad
    \cosh\theta = \frac{R_1^2 + R_2^2 - d^2}{2 R_1 R_2}
\end{equation}
We imply that the length of cylinders is equal to unity.
The real dependence is presented in Fig.~\ref{fig:forces:total}, \ref{fig:forces:inverse}.

\subsection{Gas between two elliptical cylinders}

\begin{figure}
    \centering
    \includegraphics{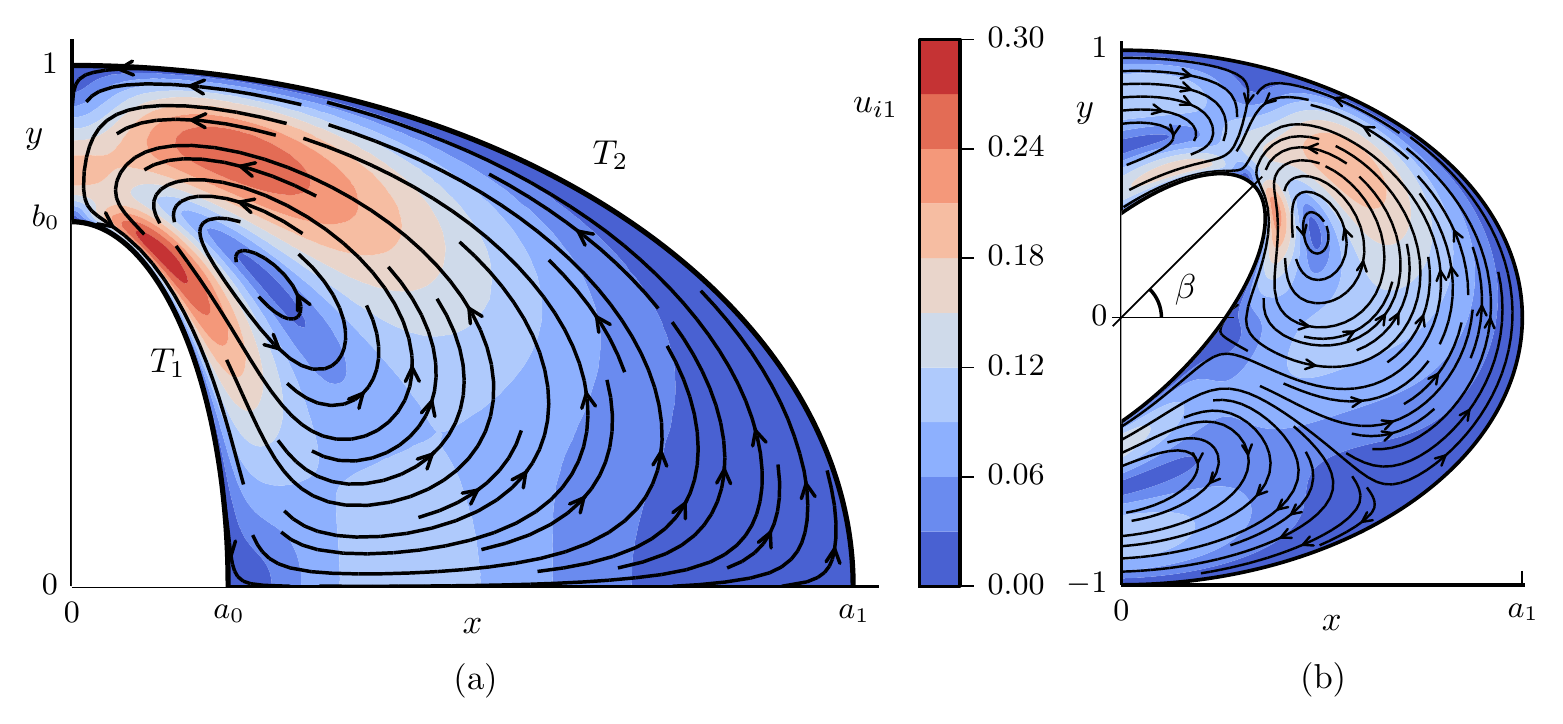}
    \caption{The velocity field \(u_{i1}\) between coaxial elliptical cylinders
            for \(\beta = \pi/2\) (a) and \(\beta = \pi/4\) (b).
        Curves with arrows indicate the direction, contour lines show the magnitude.}
    \label{fig:elliptic}
\end{figure}

In the last example, a gas is placed between two coaxial elliptical cylinders
in such a way that the major axes of the ellipses are rotated by an angle \(\beta\) in a cross section.
Let the semi-minor axis of the outer cylinder be the characteristic length and lies on the \(y\) axis,
while the semi-major one is \(a_1\) and lies on the \(x\) axis.
The semi-axes of the inner ellipse are \(a_0\) and \(b_0\) in length.
The temperature of the inner cylinder \(T_1 = 1\) and the outer one \(T_2 = 1+\tau\) again.

In Fig.~\ref{fig:elliptic} the velocity field is shown for
\[ a_1 = 1.5, \quad a_0 = 0.3, \quad b_0 = 0.7, \quad \tau = 4. \]
Numerical simulation of a rarefied gas using the DSMC method in the same geometry
for a wide range of Knudsen numbers \(0.1\le\Kn\le5\) can be found in~\cite{SoneCoaxial}.

\begin{figure}[ht]
    \centering
    \begin{minipage}{.48\textwidth}
        \centering
        \includegraphics{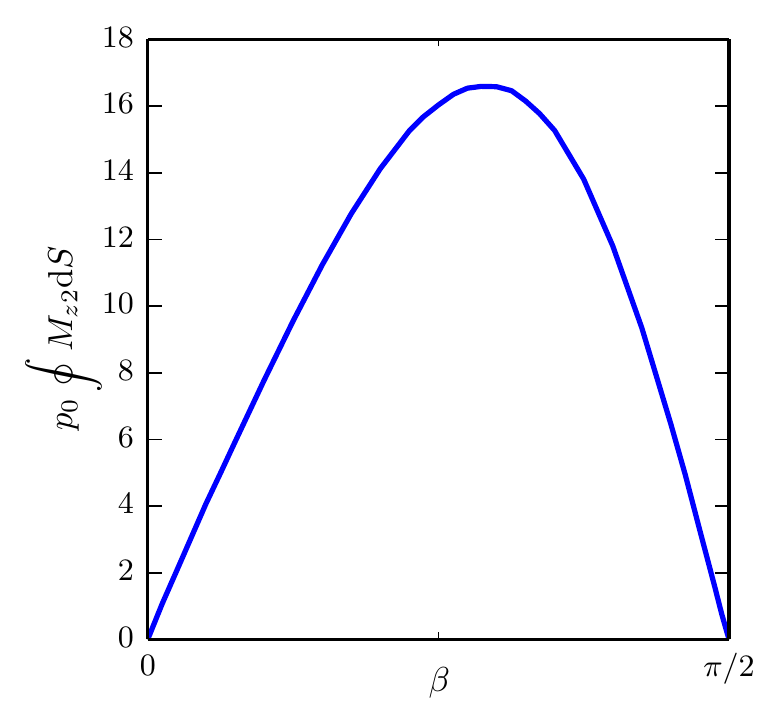}
        \caption{The total moment of force acting on the inner elliptic cylinder
                versus~\(\beta\), the angle between the major axes of the cylinders,
                for \(\tau=4\).}
        \label{fig:elliptic:beta}
    \end{minipage}
    \quad
    \begin{minipage}{.48\textwidth}
        \centering
        \includegraphics{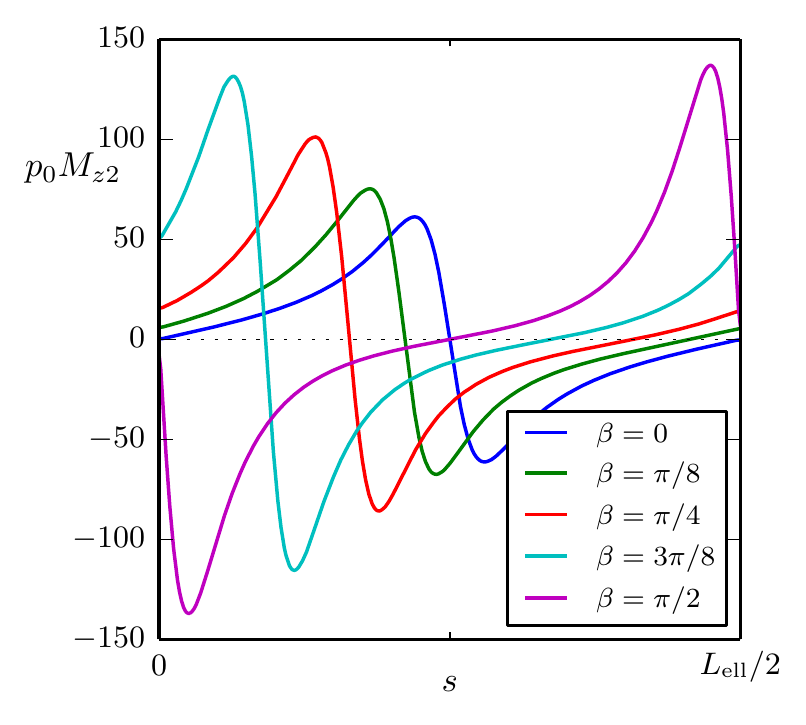}
        \caption{The profile of the sum of the components of the moment \(M_{z2}\)
                along the inner ellipse for \(\tau=4\).
                \(s=0\) and \(s=L_\mathrm{ell}/2\) correspond, respectively,
                to the upper (\(y>0\)) and lower (\(y<0\)) points on the axis of ordinates (\(x=0\)).
                \(L_\mathrm{ell}\) is the perimeter of the inner ellipse.}
        \label{fig:elliptic:profiles}
    \end{minipage}
\end{figure}

Examine the moment \(M_{i2} = e_{ijk}x_jF_{k2}\) of force acting on the inner cylinder
per unit area in dependence of the rotation angle \(\beta\).
The total moment are balanced in the symmetric cases \(\beta=0\) and \(\beta=\pi/2\),
but only the perpendicular state (\(\beta=\pi/2\)) is stable (Fig.~\ref{fig:elliptic:beta}).
The corresponding profiles (in a sense described in the previous subsection)
of the moment of force along the inner ellipse for different \(\beta\)
are presented in Fig.~\ref{fig:elliptic:profiles}.

\section{Conclusion}

In the present paper, OpenFOAM\textregistered{}, a free and open-source CFD platform,
widely used and rapidly extending, is upgraded to deal with
slightly rarefied gas flows driven by significant temperature variations.
A finite-volume solver has been introduced on the basis of the appropriate
equations and boundary conditions, derived from the asymptotic analysis of the Boltzmann equation.
The program code has been validated by means of some benchmark simulations,
presented as illustrations.

Typical temperature driven flows of the first order of the Knudsen number have been considered:
thermal creep flow and nonlinear thermal-stress flow.
Forces of the second order of the Knudsen number, arising in the gas,
has been studied for several problems too.

OpenFOAM\textregistered{}, together with the newly implemented numerical solver,
is a robust tool for numerical analysis of slightly rarefied gas problems
on the kinetic basis. The principal advantage of the developed code is that
it is easily extensible as an open-source software.

A hard-sphere gas with the diffuse-reflection boundary condition is considered in this paper,
but various molecular models and boundary conditions can be used as well.
In addition, appropriate equations for gas mixtures can be also naturally implemented.

\begin{acknowledgements}
The author expresses his gratitude to Dr. Oscar Friedlander
for fruitful discussions and valuable comments.
\end{acknowledgements}

\bibliographystyle{spphys} 
\bibliography{springer}

\end{document}